\documentclass[preprint2,twoside]{hwo}

\usepackage{graphicx}

\newenvironment{my_enumerate}{
	\begin{enumerate}
		\setlength{\itemsep}{7pt}
		\setlength{\parskip}{0pt}
		\setlength{\parsep}{0pt}}{\end{enumerate}
}

\input{hwo.h}

\begin{document}

\title{\textbf{\LARGE Characterizing Transiting Exoplanet Atmospheres in the 2030s with the \emph{Hubble Space Telescope}}}

\author{
Joshua D.\, Lothringer,$^{1}$ 
Hannah R.\, Wakeford,$^{2}$
Robert C.\, Frazier,$^{3}$
Lili Alderson,$^{4}$
Munazza K.\, Alam,$^{1}$
David K., Sing,$^{5,6}$
Mei Ting Mak,$^{7,8}$
Nikole K. Lewis, $^{4}$  
L\'{i}a Corrales,$^{3}$ 
Eva-Maria Ahrer,$^{9}$ 
}
\affil{$^1$\small\it Space Telescope Science Institute, 3700 San Martin Drive, Baltimore, MD 21218, USA; \email{jlothringer@stsci.edu}}
\affil{$^2$\small\it School of Physics, University of Bristol, HH Wills Physics Laboratory, Tyndall Avenue, Bristol BS8 1TL, UK}
\affil{$^3$\small\it Department of Astronomy and Astrophysics, University of Michigan, Ann Arbor, MI, 48109, USA}
\affil{$^4$\small\it Department of Astronomy, Cornell University, 122 Sciences Drive, Ithaca, NY 14853, USA}
\affil{$^5$\small\it William H. Miller III Department of Physics and Astronomy, Johns Hopkins University, Baltimore, MD 21218, USA}
\affil{$^6$\small\it Morton K. Blaustein Department of Earth \& Planetary Sciences, Johns Hopkins University, Baltimore, MD 21218, USA}
\affil{$^7$\small\it Atmospheric, Oceanic, and Planetary Physics Department, University of Oxford, OX1 3PU, UK}
\affil{$^8$\small\it Department of Physics and Astronomy, Faculty of Environment, Science and Economy, University of Exeter, Exeter EX4 4QL, UK}
\affil{$^9$\small\it Max Planck Institute for Astronomy (MPIA), K\"{o}nigstuhl 17, 69117 Heidelberg, Germany}

\author{\footnotesize{\bf Endorsed by: Natalie H. Allen (JHU), Katherine A. Bennett (JHU), V. Abby Boehm (Cornell University), Hannah Diamond-Lowe (STScI), Arika Egan (JHU/APL), N\'estor Espinoza (STScI), Carlos Gascon (STScI), Lauren P. Miller (STScI), Elijah Mullens (Cornell University), Sarah E. Moran (STScI/University of Maryland), Lakeisha M. Ramos Rosado (JHU), Leonardo A. Dos Santos (STScI)}}

\begin{abstract}
  The \textit{Hubble Space Telescope} inaugurated the era of exoplanet atmospheric characterization. While the \textit{James Webb Space Telescope} has largely taken up the mantle of infrared atmospheric characterization, \textit{Hubble}'s unique short-wavelength capabilities remain unmatched. Recent theoretical advances in exoplanet atmospheric science combined with new observing strategies, like those offered by WFC3-UVIS/G280, have opened science cases that only \textit{Hubble} can address for the foreseeable future. In this white paper, we discuss these new windows into the atmospheres of other worlds, focusing on characterization of their hydrostatic lower atmosphere, and identify the critical capabilities necessary for future observations. We highlight three overall science cases that will depend on the continued short-wavelength capabilities of \textit{Hubble}: measuring aerosol scattering slopes, characterizing metal absorption in ultra-hot Jupiters, and understanding stellar activity with Transit Light Source effect decontamination and flare monitoring. Throughout, we highlight useful synergies between HST and JWST. \textit{This article is a response to the call for white papers by the Space Telescope Science Institute on ``Building a Roadmap for \textit{Hubble} science into the 2030s."}
  \\
  \\
\end{abstract}

\vspace{2cm}

\section{Introduction}

For over two decades, the \textit{Hubble Space Telescope} (HST) has served as a backbone in the characterization of exoplanet atmospheres. Spectroscopic observations of the transit of hot Jupiter HD~209458b with HST's Space Telescope Imaging Spectrograph (STIS) were the first to identify wavelength-dependent transit depths, the hallmark of an exoplanet atmosphere as seen in transit \citep{charbonneau:2002}. Since then, \textit{Hubble} has observed everything from metal ions escaping from the hottest known gas giant \citep{baldwin:2026} to methane absorption in a sub-Neptune planet with the same temperature as Earth \citep{benneke:2019b}. These discoveries have transformed our understanding of planetary systems, providing clues to planet formation, evolution, and habitability. 

Since its launch in 2021, the James Webb Space Telescope (JWST) has begun a new era of exoplanet characterization. While JWST's infrared capabilities generally now surpass those of HST, HST's short-wavelength capabilities remain unmatched. With JWST reaching only as blue as 0.5\,$\mu$m via the NIRSpec/PRISM mode \citep{rustamkulov:2022}, the entire blue-to-UV regime is effectively inaccessible without HST. These short-wavelengths are critical for three main reasons:

\begin{my_enumerate}
    \item Aerosols, especially photochemically generated hazes and small particle condensate clouds, scatter efficiently at short wavelengths, leading to strong, steep slopes in the transit depth towards blue wavelengths.

    \item Heavy metal species absorb strongly at short wavelengths, producing spectral features with transit depths that dwarf those seen at infrared wavelengths.

    \item Stellar inhomogeneities like starspots and faculae, which confound transit observations of small planets with JWST through the Transit Light Source (TLS) effect
    , show their strongest signals at short-wavelengths.
\end{my_enumerate}

In this white paper, we focus on the importance of addressing these science questions with HST in the years to come. While we focus on characterization of the hydrostatic atmosphere, HST also plays a critical role in characterizing atmospheric escape, of fundamental importance to exoplanet studies. 
This is described in greater detail in the white paper entitled ``The Role of the \emph{Hubble Space Telescope} in Advancing our Understanding of Atmospheric Escape in Exoplanets" (Dos Santos et al.).

\section{Key science questions}

Answering the key science questions below requires the blue-to-UV capabilities of HST to empirically answer and test our theoretical understanding.

\subsection{When do small-particle aerosols form on exoplanets and what is their nature?}

\begin{figure}[ht]
    \centering
    \includegraphics[width=1.0\linewidth]{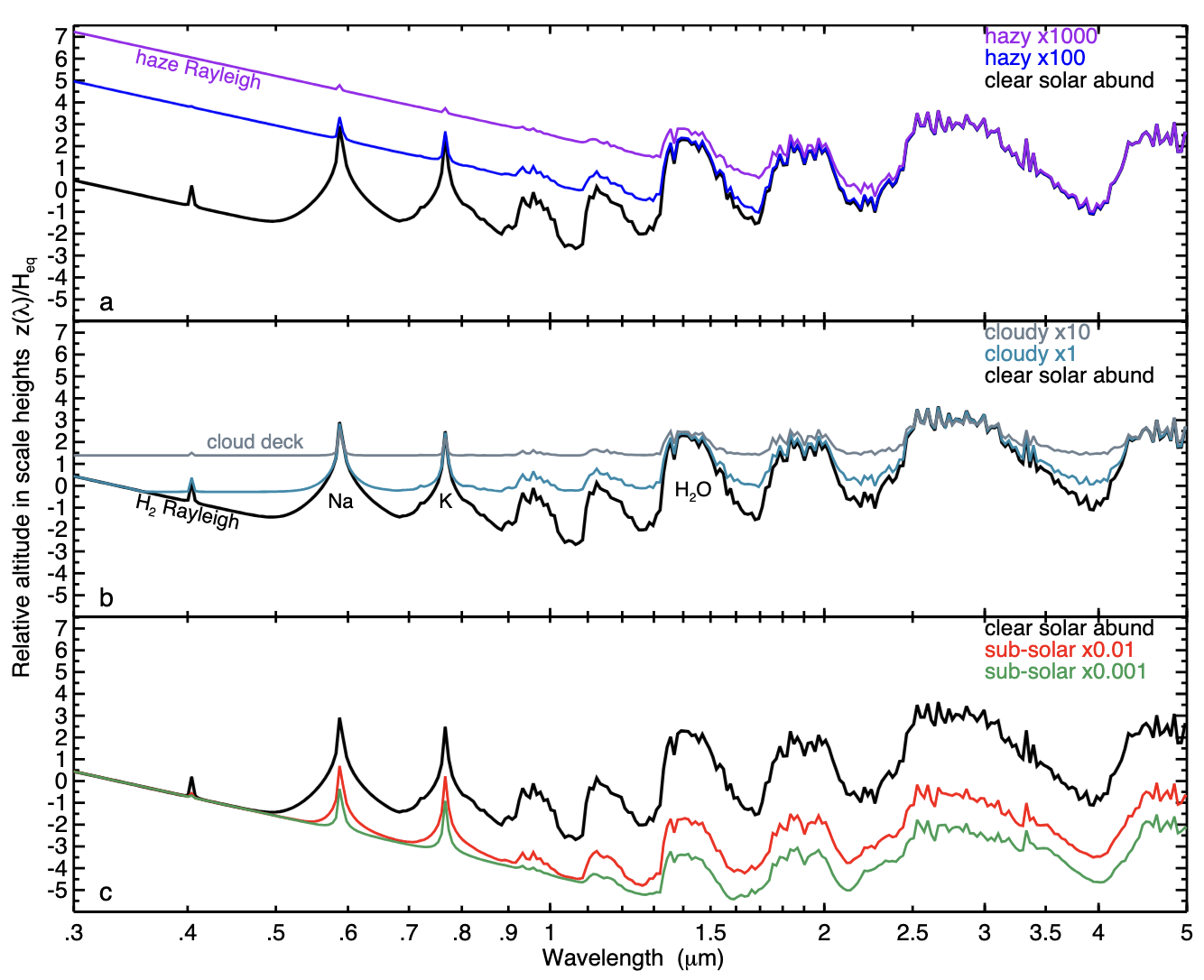}
    \caption{From \cite{sing:2016}. The effect of hazes (top), clouds (middle), and metallicity (bottom) on exoplanet transmission spectra. Short-wavelength observations ($\lesssim 0.6\,\mu$m) provide unique information.}
    \label{fig:hazes}
\end{figure}

Condensate clouds and photochemical hazes, broadly defined as aerosols, fundamentally shape the observed transmission and emission spectrum of exoplanets \citep[e.g.,][]{wakeford:2015,sing:2016,fu:2017,wakeford:2019}. Their relatively gray opacity, caused by large particles suspended high in the atmosphere, tends to mute spectral features by raising the photosphere above the clear continuum, effectively truncating spectral features \citep[e.g.,][]{gao:2020}. However, small aerosol particles can have strongly wavelength-dependent scattering at short-wavelengths due to their sub-micron sizes, creating distinctive slopes in transmission spectra \citep[see Figure~\ref{fig:hazes},][]{lecavelier:2008haze189,sing:2016,corrales:2023}. 

The small, sub-micron particles necessary to create the strong scattering slopes found in planets like HD~189733b \citep{lecavelier:2008haze189} are not predicted by most cloud formation models and so have been hypothesized to be formed via photochemical hazes \citep[e.g.,][]{he:2018a,corrales:2023}, like those seen on Titan \citep{courtin:2005}. Sub-micron particles are also necessary to create the 10\,$\mu$m silicate feature seen in brown dwarfs \citep{cushing:2006}. However, newer microphysical cloud formation models may be able to form particles small enough to create a haze-like scattering slope and strong 10$\mu$m silicate features in the MIR \citep{kiefer:2026}.

HST's short-wavelength capabilities remain crucial for this science question: only HST can constrain scattering in the NUV and in the blue-optical where it is expected to be the strongest. Recent programs (HST-GO-17162 \citep{HST-GO-17162}, HST-GO-17183 \citep{hustle:2022}) have sought to do exactly this, with data currently under analysis. While some targets become exceedingly dim at short wavelengths, programs like HST-GO-17162 are able to stack many visits together for increased precision of these targets. 

\begin{figure}[ht]
    \centering
    \includegraphics[width=1\linewidth]{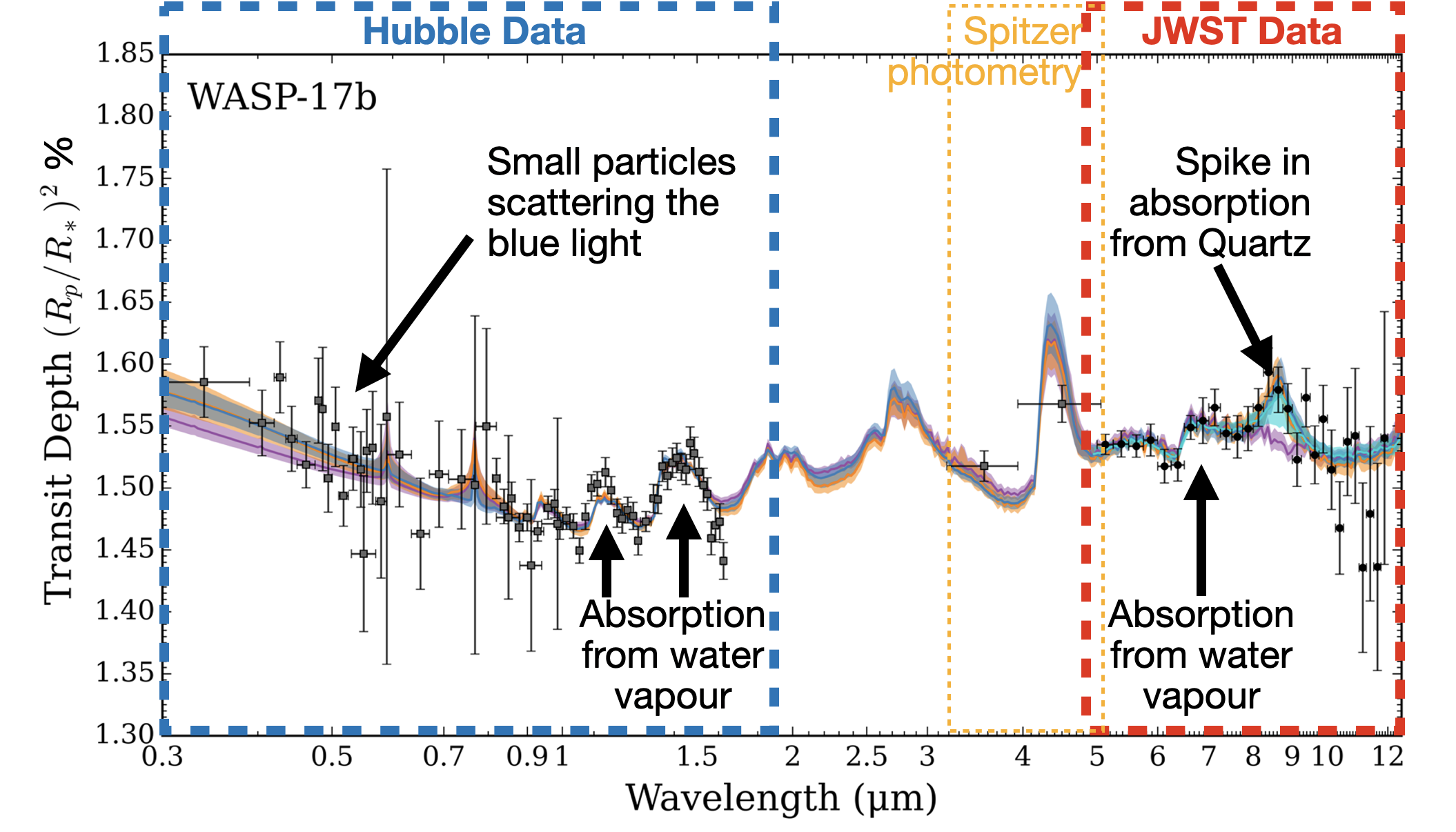}
    \caption{Modified from \citet{grant:2023}. The importance of including HST data in constraining aerosol absorption features in the MIR with JWST. Scattering in the UV-optical defines cloud particle sizes with NIR absorption defining the grey opacity defining the pressure level in the atmosphere.}
    \label{fig:quartz}
\end{figure}

A clear synergy exists between HST and JWST here. As 10\,$\mu$m silicate features are observed with JWST/MIRI/LRS \citep{grant:2023,dyrek:2024}, scattering features at short-wavelengths can then be characterized spectroscopically with HST's STIS, COS, or WFC3 instruments. In doing so, a correspondence can be drawn between the presence of scattering and MIR absorption to identify whether the two features are both caused by the same population of solid particles, or whether two distinct processes and species are responsible for each. Only with the inclusion of HST data from \citet{alderson:2022} was the first distinct aerosol feature measured with JWST able to constrain the particle sizes and pressure extent of the cloud \citep[Figure \ref{fig:quartz}, ][]{grant:2023}.  

Retrieval analyses have further shown that short wavelength observations can be critical for minimizing degeneracies in atmospheric retrievals seeking to measure bulk atmospheric properties like metallicity and C/O from JWST observations. \citet{fairman:2024} shows that the information content gained by including sequentially shorter wavelength data, beyond the reach of JWST, is vital for constraining the cloud properties of the atmosphere. Even if one does not care about clouds or hazes, short-wavelength data is highly valuable for improving the constraint of other atmospheric properties based on longer-wavelength data \citep[e.g.,][]{verma:2025}.

Another synergy between HST and JWST lies in the characterization of photochemical hazes. Perhaps the first scientific surprise of the JWST-era was the detection of photochemically-produced SO$_2$ in the atmosphere of hot Jupiter WASP-39b \citep{rustamkulov:2022,tsai:2023}. The characterization of this photochemistry has thus become a central science case for JWST observations of exoplanets. Some of these same photochemical reactions are expected to produce solid haze particles \citep[e.g.,][]{moran:2018,he:2018a} that could be responsible for scattering at short wavelengths. Joint observations with JWST (to measure photochemical products) and HST (to measure haze scattering) can shed light on the formation of these hazes and the efficacy of photochemistry.

\subsection{Which planets exhibit NUV absorption by heavy elements? What does this tell us about rainout, cloud formation, and planet formation?}

\begin{figure*}[ht]
    \centering
    \includegraphics[width=0.9\linewidth]{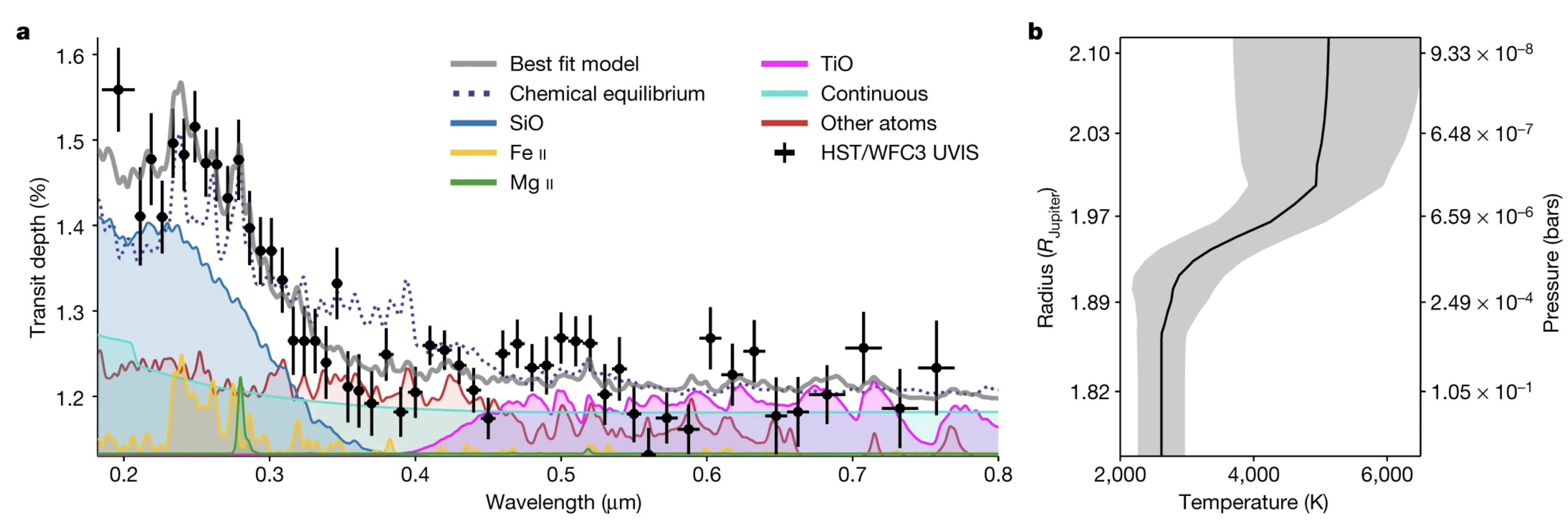}
    \caption{From \cite{lothringer:2022}. The HST/WFC3-UVIS/G280 transmission spectrum of WASP-178b shows enormous absorption in the NUV by metal species. Observations of other planets hint at a trend of NUV metal absorption increasing somewhere between equilibrium temperatures of 1900 and 2200~K.}
    \label{fig:w178}
\end{figure*}

Ultra-hot Jupiters (T$_{\mathrm{day}} \simeq$ 2200 K) represent some of the most observationally favorable targets with both JWST and HST: their characteristic clear dayside, large inflated size, high temperature, short periods, and UV-bright host stars make them ideal candidates for a variety of science cases \citep{lothringer:2018b,parmentier:2018}. It is only in these planets that we can expect to observe gaseous heavy metal species, and some only in the NUV. The presence or absence of these metals as a function of atmospheric temperature can constrain the efficacy of rainout and cold-trapping \citep{sing:2019,lothringer:2020b}, which in turn provides empirical insight into formation of condensate clouds. This also means that both refractory and volatile abundances can be measured, something that we cannot measure in the Solar System giant planets! Such abundance measurements holds highly constraining information for pinpointing exoplanet formation, migration, and evolution \citep{lothringer:2020d,chachan:2023}.


The recent advent of the utilization of HST's WFC-UVIS/G280 mode for exoplanet transit spectroscopy \citep{wakeford:2020} has opened the door to advances in the characterization of the short-wavelength spectrum of ultra-hot Jupiters; programs targeting such planets have revealed enormous absorption from metal species like Fe II, Mg II, and/or SiO extending nearly 20 scale-heights \citep[see Figure~\ref{fig:w178},][]{lothringer:2022,chachan:2025}. These species are the precursors to condensate silicate minerals, which are expected to make up the vast majority of the cloud mass in hot Jupiters and brown dwarfs \citep{wakeford:2017a}. The spectral features from these species are the largest ever measured outside of atmospheric escape, dwarfing IR spectral features measured with JWST, which typically extend only a few scale-heights in the atmosphere. Cooler planets on the hot--ultra-hot boundary appear to show much more muted NUV absorption \citep{lewis:2020,gascon:2025,gressier:2023}, hinting at a fundamental change in the behavior of giant planet atmospheres over a transition temperature range. As we move fully into the hot Jupiter regime, WFC3-UVIS/G280 can be used to reveal large-scale population trends in heavy element absorption down to aerosol formation \citep[e.g.,][]{boehm:2025}. 


We are only just beginning to piece together the trends in the behavior of these short-wavelength spectra, so the list of targets to be observed and science to be done is long. Pinpointing the exact temperatures at which different metal species evaporate into the gas phase will be highly constraining for 1D and 3D atmosphere models of cloud formation, circulation, and cold-trapping. Future observations paired with JWST data will measure the refractory-to-volatile abundance ratio with UV-to-IR spectra, as was recently done with WASP-178b \citep{lothringer:2025}, providing highly-constraining information on planet formation and migration.

\subsection{How can we disentangle the effects of stellar activity on exoplanet transit spectra? What does this activity tell us about the host star itself?}

When interpreting the observed transit spectrum of a given planet, it is often assumed that the host star was a uniform disk, modulo limb darkening. However, stellar activity in the form of star spots, faculae, and plages can contaminate the signal from the planet's atmosphere because the chord transited by the planet is no longer representative of the rest of the stellar disk \citep{oshagh:2013,mccullough:2014,rackham:2018,rackham:2019}. While occulted activity regions complicate the light curve fitting, unocculted activity regions can be insidious because one might not otherwise realize the stellar activity is affecting the planet's transmission spectrum. Stellar contamination is currently \emph{the} limiting factor for our characterization of small exoplanets with JWST, including the detection of atmospheres around rocky planets, because the signal from the planet's atmosphere can be smaller than the signal from stellar contamination \citep{rackham:2024,kreidberg:2025}.

Fortunately, techniques have been (and are being) developed to disentangle the stellar contamination signal from the planet's atmospheric signal \citep{huitson:2013,chachan:2019,wakeford:2019b,rackham:2024}. These techniques often rely on broad-wavelength coverage since the stellar contamination signal will generally increase at short wavelengths, where the spectroscopic contrast between the stellar photosphere and activity regions are largest. Thus, short-wavelength transit spectra are key to accounting for stellar contamination signals. This was recently demonstrated with JWST/NIRSpec/G395H observations of super-Earth GJ 486b, which was compatible with a water-rich atmosphere \textit{or} stellar contamination \citep{moran:2023}; only with shorter wavelength data recently obtained with JWST program 5866 \citep{JWST-GO-5866} will we be able to identify the correct atmospheric scenario.

For future JWST observations of small exoplanets, it may be especially advantageous to observe simultaneously with short-wavelength modes on HST to provide important stellar context: even if the planet's atmospheric signal is too small to measure with HST, signals from stellar contamination can still be constrained, especially at short wavelengths, as was done for Rocky Worlds DDT target LTT-1445A~b \citep{bennett:2025}. This design of contemporaneous transit coverage greatly enhances the science return of JWST's most challenging exoplanet observations and is currently being used in JWST-HST joint programs like JWST-GO-12237 (PI Boehm) and HST-GO-18250 (PI Diamond-Lowe). Indeed, this optical-to-IR coverage is exactly the methodology being explored by NASA's \textit{Pandora} SmallSat mission \citep{rackham:2026}.

\subsubsection{Flare Monitoring with HST}

A related science case for short-wavelength observations with HST also concerns understanding the behavior of the host-star, namely flare characterization \citep[e.g.,][]{loyd:2018}. HST has been the workhorse for host star characterization since the early days of exoplanet characterization, with the MUSCLES surveys being the quintessential example \citep{france:2016}. Such observations have been prerequisite inputs into models of atmospheric evolution, climate, escape, and photochemistry \citep[e.g.,][]{peacock:2019,dosSantos:2022,tsai:2023} and we recommend HST's capabilities be preserved to continue enabling such observations, especially in the UV. We also highlight a new use of HST's panchromatic capabilities to monitor for flares.

\begin{figure}[ht]
    \centering
    \includegraphics[width=1.0\linewidth]{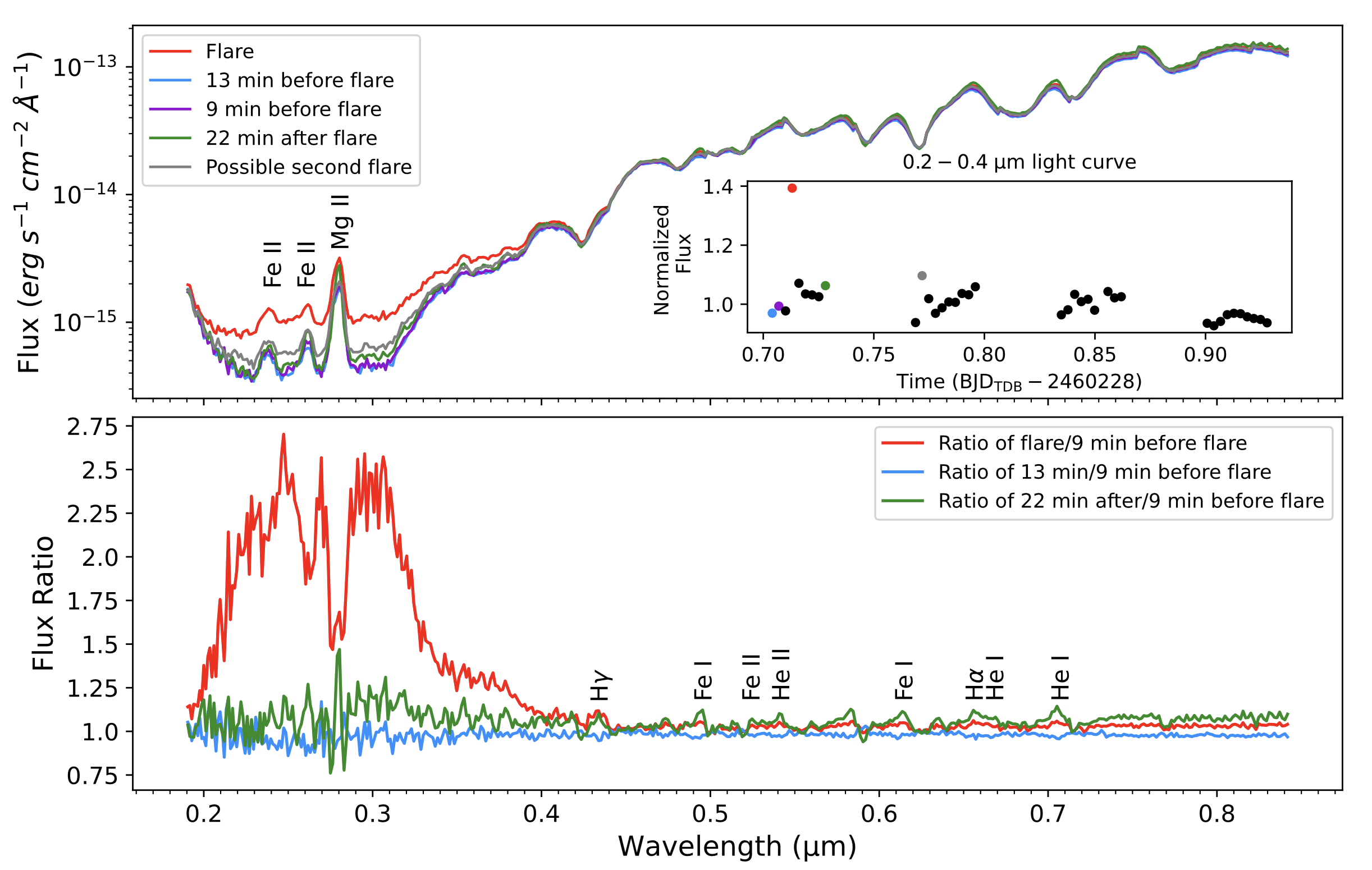}
    \caption{From \cite{bennett:2025}. The flare spectrum of LTT-1445C from time-series with HST/WFC3/G280. Current models of M-dwarf flare production cannot explain the qualitative features of the panchromatic flare spectrum.}
    \label{fig:flare}
\end{figure}

HST/WFC3/G280 transit time-series of LTT-1445Ab was originally obtained with HST Program 16039 to characterize the atmosphere of one of the most observationally favorable super-Earth planets. However, during the observations, a massive flare of one of the binary M-dwarf companion stars, LTT-1445C, was observed \citep{bennett:2025}. Because of the wide UV-to-optical wavelength coverage, the flare was characterized in unprecedented detail, confounding current models of flare production (see Figure~\ref{fig:flare}). Follow-up observations in HST Program 18140 will monitor Proxima Centauri over 28 orbits with a similar set-up. Characterization of current and future exoplanet host star targets may benefit from similar observations and capabilities in the years to come.

\section{Relevant \emph{HST} observing modes and capabilities}

Nearly all of HST's NUV spectroscopy modes have been used to characterize exoplanet atmospheres in one way or the other over the past two decades. At high spectral resolution, STIS's echelle modes can measure escaping metal ions \citep{vidal-madjar:2004,sing:2019,cubillos:2023,lothringer:2022,baldwin:2026}, though the precision is generally too low to measure, except when heavily binned with respect to wavelength \citep{sing:2019,gressier:2023}. A wide array of lower resolution STIS and COS FUV and NUV modes have been used to measure atmospheric escape of hydrogen, volatiles, and metals \citep[e.g.,][]{ehrenreich:2015,fossati:2010}. 

For constraints on the hydrostatic photosphere-level atmosphere, modes like STIS/G430L and G750L were originally used to measure the full optical transit spectrum \citep[e.g.,][]{sing:2008b,sing:2016}, but WFC3-UVIS's G280 slitless grism spectroscopy mode was later recognized for its much better throughput and wider wavelength coverage (0.2--0.8~$\mu$m), at the cost of lower spectral resolution and observational complexity due to the curved and overlapping higher-order spectral traces \citep{wakeford:2020}.

Looking towards the future, the combination of STIS's high-resolution spectroscopic modes and WFC3's low-resolution G280 mode provide powerful constraints on both the hydrostatic and escaping atmospheres of giant exoplanets, as was done for observations of WASP-178b \citep[see Figure~\ref{fig:w178}][]{lothringer:2022}. Support and calibration for the variety of HST's UV-optical capabilities maximizes the power of the telescope for exoplanet science.

\section{Supporting the mission of the \emph{Habitable Worlds Observatory}}
The \textit{Habitable Worlds Observatory} (HWO), planned to observe from the UV into the IR, will be critical in finishing the scientific journey that previous flagship observatories such as \textit{Hubble}, \textit{Spitzer} and \textit{JWST} have started \citep[see][]{wakeford:2026}. While the ESA Ariel Mission has the potential to build a foundation of knowledge on optical photometric phase curves through three photometric bands that cover 0.5\,--\,0.6\,$\mu$m, 0.6\,--\,0.8\,$\mu$m, and 0.8\,--\,1.1\,$\mu$m these bands lack the needed spectral resolution to resolve critical UV species. \textit{Hubble} with its broad spectroscopic coverage from the UV-optical with both WFC3-UVIS and STIS instruments is therefore a vital bridge to this new flagship. Extended access to UV spectroscopy, even at the moderate signal to noise provided by HST, is key to our continued knowledge of exoplanet atmospheres. The three key areas of research we described are ongoing challenges in exoplanet research, with large programs such as HUSTLE (GO-17183, \citealt{hustle:2022}) and SPACE (GO-17192, \citealt{SPACE:program}) often taking many years to complete with repeated observations of the same target or covering a wide number of targets. As we move to the lower signal-to-noise regime the more observations are needed in combination to measure these vital atmospheric properties. Long term investment in HST will better bridge the gap to HWO and continue to enable ground-breaking science.


\section{Future outlook}

Any extension in \textit{Hubble}'s lifetime would be well put to use by the astronomical community as indicated by the highly competitive proposal selection rate. Over the past 25 years, encompassing the entirety of humanity's exploration of these new worlds, HST has played a central role. The unmatched UV and optical spectroscopic capabilities of \textit{Hubble} will not be replaced until the launch of HWO. 

Continued support for observational setup, calibration, scheduling (including simultaneous observations with JWST), reduction, and analysis via instrument team experts, as well as exoplanet-specific tools like EXO.MAST\footnote{\footnotesize  \url{https://exo.mast.stsci.edu/}} and the Exoplanet Characterization Toolkit (ExoCTK)\footnote{\footnotesize \url{https://exoctk.stsci.edu/}}, is vital for scientists to continue to utilize \textit{Hubble}'s capabilities to the fullest extent. \textit{Hubble}'s rich archive will also continue to be a source of new discovery in the decades to come, so support and enhancement of the Mikulski Archive for Space Telescopes (MAST) and the related TrExoLIST\footnote{\footnotesize \url{https://www.stsci.edu/~WFC3/trexolists/trexolists.html}} is similarly important.


\bibliography{ms_new}

\end{document}